\documentclass[prc,superscriptaddress,noshowpacs,unsortedaddress,twocolumn,showpacs,preprintnumbers,amsmath,amssymb]{revtex4-1}

\usepackage[dvipdfmx]{graphicx}
\usepackage{amsmath,amssymb,times}
\usepackage{color}
\usepackage{ulem}
\usepackage{bm}
\usepackage{here}

\newcommand{\beq}{\begin{equation}}
\newcommand{\eeq}{\end{equation}}
\newcommand{\bea}{\begin{eqnarray}}
\newcommand{\eea}{\end{eqnarray}}

\newcommand{\bfi}[1]{\mbox{\boldmath $#1$}}

\newcommand{\vK}{{\bfi K}}

\newcommand{\vs}{{\bfi s}}

\newcommand{\vrr}{{\bfi r}}
\newcommand{\vR}{{\bfi R}}

\begin{document}
\title{Folding-model approach to reaction cross section of $^{4,6,8}$He+$^{12}$C scattering
at 790~MeV}

\author{Shingo~Tagami}
\affiliation{Department of Physics, Kyushu University, Fukuoka 819-0395, Japan}

\author{Tomotsugu~Wakasa}
\affiliation{Department of Physics, Kyushu University, Fukuoka 819-0395, Japan}

\author{Maya~Takechi}
\affiliation{Niigata University, Niigata 950-2181, Japan}

\author{Jun~Matsui}
\affiliation{Department of Physics, Kyushu University, Fukuoka 819-0395, Japan}

\author{Masanobu Yahiro}
\email[]{orion093g@gmail.com}
\affiliation{Department of Physics, Kyushu University, Fukuoka 819-0395, Japan}             

\date{\today}

\begin{abstract}
\noindent 
Tanihata {\it et al.} determined matter radii $r_{m}(\sigma_{\rm I})$ for $^{4,6,8}$He 
 from interaction cross sections $\sigma_{\rm I}$ of  $^{4,6,8}$He+$^{12}$C scattering at 
 790~MeV per nucleon, using the optical limit of the Glauber model. 
Lu {\it et al.} determined proton radii $r_{p}({\rm AIS})$ for $^{4,6,8}$He with 
the atomic isotope shifts (AIS). 
We  investigate whether the Love-Franey $t$-matrix folding model is good for 
$^{4,6,8}$He+$^{12}$C scattering at 790~MeV per nucleon. 
\end{abstract}

\maketitle


\section{Introduction and conclusion}
\label{Introduction}

{\it Background:}
The matter radius $r_{m}$, the neutron skin $r_{\rm skin}$ and halo structure 
are important properties of nuclei. 
When a nucleus has one or more loosely-bound nucleons surrounding a tightly bound core, 
it is considered that the nucleus has a halo structure. Eventually, we may consider that   
$^{6,8}$He have the halo structure.  

Lu {\it et al.} measured the atomic isotope shifts (AIS) along $^{4,6,8}$He by performing laser spectroscopy 
on individual trapped atoms and determined proton radii $r_{p}({\rm AIS})=1.462(6), 1.934(9), 1.881(17)$~fm for 
$^{4,6,8}$He~\cite{Lu:2013ena}. 

For He isotopes, Tanihata {\it et al.} determined $r_{m}$ from interaction cross sections $\sigma_{\rm I}$~\cite{Tanihata:1988ub,Ozawa:2001hb};  $r_{m}(\sigma_{\rm I})=1.57(4), 2.48(3), 2.52(3)$~fm for $^{4,6,8}$He by assuming the the harmonic-oscillator distribution and using the optical limit of the Glauber model~\cite{Glauber}; see 
Eq. (4.8) of Ref.~\cite{Ozawa:2001hb}.  
The Glauber model is constructed with the adiabatic and the eikonal approximation in the frame work of 
many-body Schr\"{o}dinger equation~\cite{Glauber,Yahiro-Glauber}.

In Ref.~\cite{Ozawa:2001hb}, the experimental values of $r_{m}$ are accumulated  from $^{4}$He to $^{32}$Mg. 
For  $^{4,6,8}$He+$^{12}$C and $^{12}$C+$^{12}$C scattering, the  data on interaction 
cross section $\sigma_{\rm I}({\rm exp})$ are presented at 790~MeV with high accuracy of 1\%~\cite{Ozawa:2001hb}.

{\it Aim:} 
We investigate reliability of  the folding model based on Love-Franey (LF) $t$-matrix~\cite{LF} for  
$^{4,6,8}$He+$^{12}$C scattering at 790~MeV per nucleon. 

{\it Results:}
First, we apply the LF $t$-matrix folding model for $^{12}$C+$^{12}$C scattering 
at $E_{\rm lab}=790$~MeV per nucleon, and determine the ratio $F=\sigma_R({\rm LF})/\sigma_I({\rm exp})$
at  $E_{\rm lab}=790$~MeV per nucleon.  The resulting value is $F=0.9356$. 
Secondly, for $^{6,8}$He, we deduce neutron radius $r_{n}(\sigma_{\rm I})=2.71,  2.70$~fm  from 
the $r_{m}(\sigma_{\rm I})=2.48,2.52$~fm and the $r_{p}({\rm AIS})=1.934, 1.881$~fm. 
For  $^{4}$He,  we assume $r_{n}({\rm AIS})=r_{p}({\rm AIS})$, leading to  
$r_{m}({\rm AIS})=r_{n}({\rm AIS})=r_{p}({\rm AIS})$. 
Thirdly, we use Gogny-D1S HFB (GHFB), and 
scale the proton and neutron GHFB densities so as to satisfy $r_{p}({\rm scaling})=r_{p}({\rm AIS})$ and 
$r_{n}({\rm scaling})=r_{n}({\rm AIS})$ for $^{4}$He 
and  $r_{p}({\rm scaling})=r_{p}({\rm AIS})$ and $r_{n}({\rm scaling})=r_{n}(\sigma_{\rm I})$ for $^{6,8}$He.  
For $^{4,6,8}$He+$^{12}$C scattering, the reaction cross section $\sigma_{\rm R}({\rm scaling})$ calculated 
with the scaled densities reproduce the $\sigma_{\rm I}({\rm exp})$.

{\it Conclusion:} 
Our conclusion is that the Love-Franey (LF) $t$-matrix folding model  is good for $^{4,6,8}$He+$^{12}$C scattering at 790~MeV per nucleon.

\section{Model}
\label{Sec-Framework}

In the $t$-matrix folding model, the potential $U$ consists 
of the direct part ($U^{\rm DR}$) and the exchange part ($U^{\rm EX}$) 
defined by~\cite{DFM-standard-form,DFM-standard-form-2}
\bea
\label{eq:UD}
U^{\rm DR}(\vR) \hspace*{-0.15cm} &=& \hspace*{-0.15cm} 
\sum_{\mu,\nu}\int \rho^{\mu}_{\rm P}(\vrr_{\rm P}) 
            \rho^{\nu}_{\rm T}(\vrr_{\rm T})
            t^{\rm DR}_{\mu\nu}(s) d \vrr_{\rm P} d \vrr_{\rm T}, \\
\label{eq:UEX}
U^{\rm EX}(\vR) \hspace*{-0.15cm} &=& \hspace*{-0.15cm}\sum_{\mu,\nu} 
\int \rho^{\mu}_{\rm P}(\vrr_{\rm P},\vrr_{\rm P}-\vs)
\rho^{\nu}_{\rm T}(\vrr_{\rm T},\vrr_{\rm T}+\vs) \nonumber \\
            &&~~\hspace*{-0.5cm}\times t^{\rm EX}_{\mu\nu}(s) \exp{[-i\vK(\vR) \cdot \vs/M]}
            d \vrr_{\rm P} d \vrr_{\rm T},~~~~
            \label{U-EX}
\eea
where $\vs=\vrr_{\rm P}-\vrr_{\rm T}+\vR$ 
for the coordinate $\vR$ between a projectile (P) and a target (T). The coordinate 
$\vrr_{\rm P}$ 
($\vrr_{\rm T}$) denotes the location for the interacting nucleon 
measured from the center-of-mass of P (T). 
Each of $\mu$ and $\nu$ stands for the $z$-component
of isospin; 1/2 means neutron and $-$1/2 does proton.
The original form of $U^{\rm EX}$ is a non-local function of $\vR$,
but  it has been localized in Eq.~\eqref{U-EX}
with the local semi-classical approximation~\cite{Brieva-Rook} in which
P is assumed to propagate as a plane wave with
the local momentum $\hbar \vK(\vR)$ within a short range of the 
nucleon-nucleon interaction, where $M=A A_{\rm T}/(A +A_{\rm T})$
for the mass number $A$ ($A_{\rm T}$) of P (T).
The validity of this localization is shown in Ref.~\cite{Minomo:2009ds}.

The direct and exchange parts, $t^{\rm DR}_{\mu\nu}$ and 
$t^{\rm EX}_{\mu\nu}$, of the $t$ matrix are described by
\begin{align}
t_{\mu\nu}^{\rm DR}(s) 
&=
\displaystyle{\frac{1}{4} \sum_S} \hat{S}^2 t_{\mu\nu}^{S1}
 (s) \hspace*{0.1cm}  \hspace*{0.1cm} 
 {\rm for} \hspace*{0.1cm} \mu+\nu = \pm 1,
 \\
t_{\mu\nu}^{\rm DR}(s) 
&=
\displaystyle{\frac{1}{8} \sum_{S,T}} 
\hat{S}^2 t_{\mu\nu}^{ST}(s) 
\hspace*{0.1cm}  \hspace*{0.1cm} 
{\rm for} \hspace*{0.1cm} \mu+\nu = 0,
\\
t_{\mu\nu}^{\rm EX}(s) 
&=
\displaystyle{\frac{1}{4} \sum_S} (-1)^{S+1} 
\hat{S}^2 t_{\mu\nu}^{S1} (s) 
\hspace*{0.1cm}  \hspace*{0.1cm} 
{\rm for} \hspace*{0.1cm} \mu+\nu = \pm 1, 
\\
t_{\mu\nu}^{\rm EX}(s) 
&=
\displaystyle{\frac{1}{8} \sum_{S,T}} (-1)^{S+T} 
\hat{S}^2 t_{\mu\nu}^{ST}(s) 
\hspace*{0.1cm}  \hspace*{0.1cm}
{\rm for} \hspace*{0.1cm} \mu+\nu = 0 
,
\end{align}
where $\hat{S} = {\sqrt {2S+1}}$ and $t_{\mu\nu}^{ST}$ are 
the spin-isospin components of the $t$-matrix interaction.
As for the $t$-matrix interaction, we take the LF version. 
We apply the LF $t$-matrix  folding model for  $^{12}$C+$^{12}$C and $^{4,6,8}$He+$^{12}$C scattering 
at $E_{\rm lab}=790$~MeV per nucleon.

As  proton and  neutron densities, we use Gogny D1S HFB (GHFB) and scale the GHFB densities; 
see the center-of-mass (cm) corrections for Ref.~\cite{Sumi:2012fr}. 
In order to make the scaling, for $^{6,8}$He, we determine $r_{n}(\sigma_{\rm I})=2.71,  2.70$~fm  from 
the $r_{m}(\sigma_{\rm I})=2.48,2.52$~fm and the $r_{p}({\rm AIS})=1.934, 1.881$~fm. 
For  $^{4}$He,  we assume $r_{n}({\rm AIS})=r_{p}({\rm AIS})$, leading to 
$r_{m}({\rm AIS})=r_{n}({\rm AIS})=r_{p}({\rm AIS})$. Next, 
we scale the GHFB proton and neutron densities so as to $r_{p}({\rm scaling})=r_{p}({\rm AIS})$ and $r_{n}({\rm scaling})=r_{n}(\sigma_{\rm I})$ for $^{6,8}$He and $r_{p}({\rm scaling})=r_{p}({\rm AIS})$ and 
$r_{n}({\rm scaling})=r_{n}({\rm AIS})$ for $^{4}$He. 
The scaled densities only are used in the folding model. 
 
Also for $^{12}$C, we construct the densities with GHFB with the angular momentum projection 
and scale the densities so as to $r_{m}(\sigma_{\rm I}) = 2.35$~fm~\cite{Tanihata:1988ub} and the 
$r_{p}({\rm exp})=2.33$~fm~\cite{Angeli:2013epw} of electron scattering. 

\section{Results}
\label{Results} 

Firstly, we apply the LF $t$-matrix folding model for $^{12}$C+$^{12}$C scattering 
at $E_{\rm lab}=790$~MeV per nucleon, and determine the ratio
$F=\sigma_R({\rm LF})/\sigma_I({\rm exp})$ as $F=0.9356$, 
where the $\sigma_I({\rm exp})=856$~mb was measured by Tanihata {\it el al.}~\cite{Ozawa:2001hb}.

\begin{figure}[H]
\begin{center}
 \includegraphics[width=0.5\textwidth,clip]{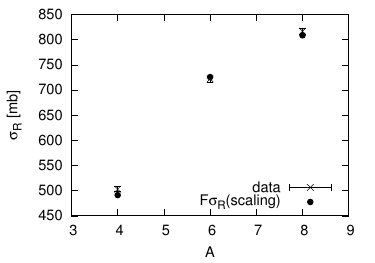}

 \caption{ 
Reaction cross sections $\sigma_{\rm R}$  for $^{4,6,8}$He+$^{12}$C scattering at 790~MeV. 
Closed circles denote results  $F \sigma_{\rm R}({\rm scaling})$. 
 The data (crosses) on $\sigma_{\rm I}$ are taken from Ref.~\cite{Ozawa:2001hb}.
   }
 \label{Fig-RXsec-C+He}
\end{center}
\end{figure}

Figure~ \ref{Fig-RXsec-C+He} shows  reaction cross sections $\sigma_{\rm R}$ 
as a function of mass number $A$ for $^{4,6,8}$He+$^{12}$C+$^{4,6,8}$He scattering at 790~MeV. 
The reaction cross sections $F \sigma_{\rm R}({\rm scaling})$ obtained with the scaled densities  
reproduce $\sigma_{\rm I}({\rm exp})$~\cite{Ozawa:2001hb}; 
note that the data have high accuracy of 1~\%. 
This success indicates that the LF $t$-matrix folding model is good 
for  $^{4,6,8}$He+$^{12}$C scattering at 790~MeV per nucleon.

We determine $r_{m}({\rm exp})$ from the $\sigma_{\rm I}({\rm exp})$; the results  
are $r_{m}({\rm exp})=1.462(6), 2.48(3), 2.53(3)$~fm for $^{4,6,8}$He.  
Our results are close to $r_{m}(\sigma_{\rm I})=1.57(4)$~fm~\cite{Tanihata:1988ub} for $^{4}$He and agree with 
$r_{m}(\sigma_{\rm I})=2.48(3), 2.52(3)$~fm~\cite{Tanihata:1988ub} for $^{6,8}$He.

\noindent
\appendix

\noindent
\begin{acknowledgments}
We would like to thank Dr. Toyokawa for providing his code. 
\end{acknowledgments}



\bibliographystyle{prsty}

\end{document}